\documentstyle[preprint,aps,epsfig,tighten]{revtex}

\renewcommand{\deg}{^{\circ}}

\newcommand{\pip}[1]{\left( #1 \right)} 
\newcommand{\pib}[1]{\left[ #1 \right]}
\newcommand{\ov}[1]{\overline {#1}}

\newcommand{\wsep}[2]{\hspace{#1 in} \mbox{#2} \hspace{#1in}}

\newcommand{\w}[1]{\mbox{#1}}

\newcommand{\eqnb}{\begin{equation}}
\newcommand{\eqne}{\end{equation}}

\newcommand{\question}[1]{}
\newcommand{\NSt}{{\mbox{\scriptsize\it NS}}}
\newcommand{\St}{{\mbox{\scriptsize\it S}}}

\begin{document}

\draft

\title{On the \mbox{$\eta$--$\eta^\prime$} complex in the SD--BS approach}

\author{D. Klabu\v{c}ar$^a$, D.\ Kekez$^b$, M. D. Scadron$^c$}
\address{{\footnotesize $^a$Physics Department, Faculty of Science, 
University of Zagreb,} \\ {\footnotesize Bijeni\v{c}ka c. 32, 
Zagreb 10000, Croatia}}

\address{{\footnotesize $^b$Rudjer Bo\v{s}kovi\'{c} Institute,
  P.O.B. 180, 10002 Zagreb, Croatia}}

\address{{\footnotesize $^c$Physics Department, University of Arizona,
  Tucson, AZ 85721, USA}}

\maketitle


\begin{abstract}
\noindent The bound-state Schwinger-Dyson and Bethe-Salpeter (SD--BS) 
approach is chirally well-behaved and provides a reliable treatment of 
the $\eta$--$\eta^\prime$ complex although a ladder approximation is
employed. Allowing for the effects of the SU(3) flavor symmetry breaking 
in the quark--antiquark annihilation, leads to the improved 
$\eta$--$\eta^\prime$ mass matrix.

\end{abstract}

\section{\mbox{$\eta$--$\eta^\prime$} phenomenology and Goldstone structure}
\label{Preliminaries}

The physical isoscalar pseudoscalars $\eta$ and $\eta'$ are usually given as
\begin{equation}
|\eta\rangle = \cos\theta\, |\eta_8\rangle
                       - \sin\theta\, |\eta_0\rangle~,
\,\,\,\,\,\,\,
|\eta^\prime\rangle = \sin\theta\, |\eta_8\rangle
                              + \cos\theta\, |\eta_0\rangle~,
\label{eta-etaPrimeDEF}
\end{equation}
{\it i.e.}, as the orthogonal mixture
of the respective octet and singlet isospin zero states, $\eta_8$ and
$\eta_0$. In the flavor SU(3) quark model, they are defined through
quark--antiquark ($q\bar q$) basis states $|f\bar{f}\rangle$ ($f=u,d,s$)
as
\begin{mathletters}
\label{eta8-eta0def}
        \begin{eqnarray}
        |\eta_8\rangle
        &=&
        \frac{1}{\sqrt{6}} (|u\bar{u}\rangle + |d\bar{d}\rangle
                                            -2 |s\bar{s}\rangle)~,
\label{eta8def}
        \\
        |\eta_0\rangle
        &=&
        \frac{1}{\sqrt{3}} (|u\bar{u}\rangle + |d\bar{d}\rangle
                                             + |s\bar{s}\rangle)~.
\label{eta0def}
        \end{eqnarray}
\end{mathletters}
The exact SU(3) flavor symmetry ($m_u = m_d = m_s$) is nevertheless
badly broken. It is an excellent approximation to assume the exact
isospin symmetry ($m_u = m_d$), and a good approximation to take even
the chiral symmetry limit for $u$ and $d$-quark
(where {\it current} quark masses $m_u = m_d = 0$), but
for a realistic description, the strange quark mass $m_s$ must be
significantly heavier than $m_u$ and $m_d$. The same holds for the
{\it constituent} quark masses, denoted by $\hat{m}$ for {\it both} 
$u$ and $d$ quarks since we rely on the isosymmetric limit, and by 
$\hat{m}_s$ for the $s$-quark. They are nonvanishing in the chiral 
limit (CL). In the strange sector, CL is useful only qualitatively, 
as a theoretical limit. (CL would reduce $\hat{m}_s$ to $\hat{m}$, 
on which CL has almost negligible influence.)

Thus, with $|u\bar{u}\rangle$ and $|d\bar{d}\rangle$ being practically
chiral states as opposed to a significantly heavier $|s\bar{s}\rangle$,
Eqs.~(\ref{eta8-eta0def}) do not define the octet and singlet states 
of the exact SU(3) flavor symmetry, but the {\it effective} octet and 
singlet states. Hence, as in Ref. \cite{KlKe2} for example, only in
the sense that the same $q\bar q$ states $|f\bar{f}\rangle$ ($f=u,d,s$)
appear in both Eq.~(\ref{eta8def}) and Eq.~(\ref{eta0def}) do these
equations implicitly assume nonet symmetry (as pointed out by Gilman and 
Kauffman \cite{GilKauf}, following Chanowitz, their Ref.~[8]). However, 
in order to avoid the U$_A$(1) problem, this symmetry must ultimately be
broken at least at the level of the masses. In particular, it must
be broken in such a way that $\eta \to \eta_8$ becomes massless but
$\eta' \to \eta_0$ remains massive (as in Ref. \cite{KlKe2}) when
CL is taken for all three flavors, $m_u, m_d, m_s \to 0$.
Nevertheless, the CL-vanishing octet eta mass $m_{\eta_8}$ is rather 
heavy for the realistically broken SU(3) flavor symmetry;
for the empirical pion and kaon masses $m_\pi$ and $m_K$,
the Gell-Mann-Okubo mass formula $m_\pi^2 + 3m_{\eta_{8}}^2 = 4m_K^2$
yields $m_{\eta_8} \approx 567$ MeV. In that case, and for the 
empirical masses of $\eta (547)$ and $\eta'(958)$, the singlet 
$\eta_0$ mass $m_{\eta_0}$ (nonvanishing even in CL) can 
be found from the mass--matrix trace 
\begin{equation}
        m_{\eta_8}^2 + m_{\eta_0}^2 =
        m_{\eta}^2 + m_{\eta'}^2 \approx
        1.22 ~\w{GeV}^2,\wsep{.1}{giving}
        m_{\eta_0} \approx 947 ~\w{MeV} \, .
        \label{eqno1}
\end{equation}

Alternatively, one can work in a nonstrange ({\it NS})--strange ({\it S})
basis:
\begin{mathletters}
\label{NS-Sbasis}
        \begin{eqnarray}
        |\eta_\NSt\rangle
        &=&
        \frac{1}{\sqrt{2}} (|u\bar{u}\rangle + |d\bar{d}\rangle)
  = \frac{1}{\sqrt{3}} |\eta_8\rangle + \sqrt{\frac{2}{3}} |\eta_0\rangle~,
\label{etaNSdef}
        \\
        |\eta_\St\rangle
        &=&
            |s\bar{s}\rangle
  = - \sqrt{\frac{2}{3}} |\eta_8\rangle + \frac{1}{\sqrt{3}} |\eta_0\rangle~.
\label{etaSdef}
        \end{eqnarray}
\end{mathletters}
In analogy with Eq. (\ref{eqno1}), in this basis one finds
\begin{equation}
	m_{\eta_{NS}}^2 + m_{\eta_S}^2 = 
	m_\eta^2 + m_{\eta'}^2 \approx
	1.22 ~\w{GeV}^2,
	\label{eqno2}
\end{equation}
whereas the {\it NS--S} mixing relations, diagonalizing the mass matrix, are
\begin{equation}
|\eta\rangle = \cos\phi_P |\eta_\NSt\rangle
             - \sin\phi_P |\eta_\St\rangle~,
\,\,\,\,\,\,\,
|\eta^\prime\rangle = \sin\phi_P |\eta_\NSt\rangle
             + \cos\phi_P |\eta_\St\rangle~.
\label{eqno3}
\end{equation}
The singlet-octet mixing angle $\theta$, defined by
Eqs.~(\ref{eta-etaPrimeDEF}), is related
to the {\it NS--S} mixing angle $\phi$ above as \cite{14} 
$\theta = \phi - \arctan \sqrt{2} =  \phi - 54.74\deg$.

Although mathematically equivalent to the $\eta_8$--$\eta_0$ basis, the
{\it NS--S} mixing basis is more suitable for most quark model considerations,
being more natural in practice when the symmetry between the {\it NS} and {\it S}
sectors is broken as described in the preceding passage.
There is also another important reason to keep in mind the
$|{\eta_\NSt}\rangle$-$|{\eta_S}\rangle$ state mixing angle $\phi$.
This is because it offers the quickest way to show the consistency of
our procedures and the corresponding results obtained using just one 
($\theta$ or $\phi$) {\it state} mixing angle,
with the two-mixing-angle scheme considered in Refs.
\cite{Leutwyler98,KaiserLeutwyler98,FeldmannKroll98EPJC,FeldmannKroll98PRD,FeldmannKrollStech98PRD,FeldmannKrollStech99PLB,Feldmann99IJMPA},
which is defined with respect to the mixing of the decay constants.
For clarification of the relationship with, and our results in
the two-mixing-angle scheme, we refer to Ref. \cite{KeKlSc2000}, 
particularly to its Appendix.
Here, we simply note that our considerations will ultimately lead
us to $\phi \approx 42^\circ$, practically the same as the result of 
Refs. \cite{FeldmannKroll98EPJC,FeldmannKroll98PRD,FeldmannKrollStech98PRD,Feldmann99IJMPA} and in agreement with data ({\it e.g.}, see Table 2 of Feldmann's 
review \cite{Feldmann99IJMPA}).

\begin{figure}[t]
\centerline{\psfig{figure=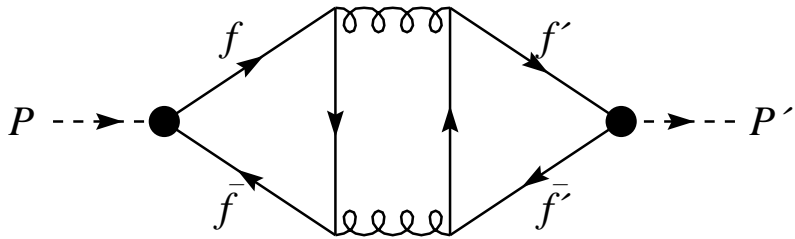,height=1.8in}}
\caption{ Nonperturbative QCD quark annihilation illustrated by the
two-gluon exchange diagram.  It shows the transition of the
$f\bar f$ pseudoscalar $P$ into the pseudoscalar $P^\prime$ having
the flavor content $f^\prime {\bar f}^\prime$. The dashed lines and
full circles depict the $q\bar q$ bound-state pseudoscalars and vertices,
respectively.
}
\end{figure}

As for a theoretical determination of the $\eta$--$\eta^\prime$ mixing 
angle $\phi$ or $\theta$, we 
follow the path of Refs.~\cite{14}. The contribution of the
gluon axial anomaly to the singlet $\eta_0$ mass is 
essentially just parameterized and not really calculated,
but some useful information can be obtained from the isoscalar
$q\bar q$ annihilation graphs of which the ``diamond" one in Fig.\ 1 
is just the simplest example. That is, we can take Fig.\ 1 in the 
nonperturbative sense, where the
two-gluon intermediate ``states'' represent any even number of
gluons when forming a C$^+$ pseudoscalar $\ov{q}q$ meson \cite{17},
and where quarks, gluons and vertices can be dressed nonperturbatively,
and possibly include gluon configurations such as instantons.
Factorization of the quark propagators in Fig.\ 1 characterized by the 
ratio $X \approx \hat{m} /\hat{m}_s$ leads to the $\eta$--$\eta^\prime$ 
mass matrix in the {\it NS--S} basis \cite{14}
\eqnb
	\pip{
		\begin{array}{ll}
	m_\pi^2 + 2 \beta	&  \, \, \,\, \quad \sqrt{2} \beta X \\
		\, \, \sqrt{2}\beta X	& \, 2 m_K^2 - m_\pi^2 + \beta X^2
		\end{array}
	}
\begin{array}{c} \vspace{-2mm} \longrightarrow \\ \phi \end{array}
	\pip{
		\begin{array}{ll}
			m_\eta^2	& 0 \\
			0		& m_{\eta'}^2
		\end{array}
	},
	\label{eqno6}
\eqne
where $\beta$ denotes the total annihilation strength of the
pseudoscalar $q\bar q$ for the {\it light} flavors $f=u,d$,
whereas it is assumed attenuated by a factor $X$ when a $s\bar s$
pseudoscalar appears. (The mass matrix in the $\eta_8$-$\eta_0$
basis reveals that in the $X\to 1$ limit, the CL-nonvanishing 
singlet $\eta_0$ mass is given by $3\beta$.)
The two parameters on the left-hand-side (LHS) of (\ref{eqno6}), $\beta$ and
$X$, are determined by the two diagonalized $\eta$ and $\eta'$
masses on the RHS of (\ref{eqno6}).  The trace and determinant of the
matrices in (\ref{eqno6}) then fix $\beta$ and $X$ to be \cite{14}
\eqnb
	\beta =
	\frac	{ (m_{\eta'}^2 - m_\pi^2) (m_\eta^2 - m_\pi^2) }
			{ 4 (m_K^2 - m_\pi^2) }
	\approx
	0.28 ~\w{GeV}^2~~,\hspace{.3in} X \approx 0.78~,
	\label{eqno7}
\eqne
with the latter value suggesting a {\it S/NS} constituent quark mass ratio
$X^{-1} \sim \hat{m}_s / \hat{m} \sim 1.3$~, near the values in 
Refs.~\cite{17,15,16,18,19}, $\hat{m}_s / \hat{m} \approx 1.45$. 

This fitted nonperturbative scale of $\beta$ in (\ref{eqno7}) depends only on
the gross features of QCD.  If instead one treats the QCD graph of
Fig.\ 1 in the perturbative sense of literally two gluons
exchanged, then one obtains \cite{20} only $\beta_{2g} \sim 0.09
~\w{GeV}^2$, which is about $1/3$ of the needed scale of $\beta$
found in (\ref{eqno7}).
(This indicates that just the perturbative ``diamond" graph 
can hardly represent even the roughest approximation to the effect of 
the gluon axial anomaly operator 
$\epsilon^{\alpha\beta\mu\nu} G^a_{\alpha\beta} G^a_{\mu\nu}$.)
The above fitted quark annihilation (nonperturbative) scale $\beta$ in
(\ref{eqno7}) can be converted to the {\it NS--S} $\eta$--$\eta^\prime$ mixing angle
$\phi$ in (\ref{eqno3}) from the alternative mixing relation 
$\tan2 \phi = 2 \sqrt{2} \beta X(m_{\eta_S}^2-m_{\eta_{NS}}^2)^{-1} 
\approx 9.02$ to \cite{14} 
\eqnb
	\phi = \arctan
	\pib{
		\frac	{(m_{\eta'}^2 - 2m_K^2 + m_\pi^2) (m_\eta^2 -
				 m_\pi^2)}
				{(2m_K^2 - m_\pi^2 - m_\eta^2) (m_{\eta'}^2
				 - m_\pi^2)}
	}^{1/2} \approx
	41.84\deg ~.
	\label{eqno8}
\eqne
This kinematical QCD mixing angle (\ref{eqno8}) or $\theta = \phi -54.74^\circ
\approx -12.9^\circ$ has dynamical analogs \cite{KlKe2,KeKlSc2000}, 
namely the coupled SD-BS approach discussed below, in Sec. \ref{MainSec}.
Since this predicted $\eta$--$\eta^\prime$ mixing angle in (\ref{eqno8}) is 
compatible with the values 
repeatedly extracted in various empirical ways
\cite{15,16}, and more recently from the
FKS scheme and theory \cite{FeldmannKroll98EPJC,FeldmannKroll98PRD,FeldmannKrollStech98PRD,FeldmannKrollStech99PLB,Feldmann99IJMPA},
we confidently use the value (\ref{eqno8}) in the mixing angle relations 
(\ref{eqno3}) to infer the nonstrange and strange $\eta$ masses,
\begin{mathletters}
\label{eqno9}
\begin{eqnarray}
	m_{\eta_{NS}}^2 &=& \cos^2 \phi ~m_\eta^2 +
	\sin^2 \phi ~m_{\eta'}^2 \approx (757.9~\w{MeV})^2
	\label{eqno9a}
\\
	m_{\eta_{S}}^2 &=& \sin^2 \phi ~m_\eta^2 + \cos^2 \phi ~m_{\eta'}^2
	\approx (801.5~\w{MeV})^2~~.
	\label{eqno9b}
\end{eqnarray}
\end{mathletters}

Thus it is clear that the true physical masses $\eta (547)$ and
$\eta' (958)$ are respectively much closer to the Nambu-Goldstone
(NG) octet $\eta_8 (567)$ and the non-NG singlet $\eta_0 (947)$
configurations than to the nonstrange $\eta_\NSt (758)$ and strange
$\eta_S (801)$ configurations inferred in Eqs.\ (\ref{eqno9}).  However, the
mean $\eta$--$\eta^\prime$ mass $(548 + 958) /2 \approx 753 ~\w{MeV}$ is
quite near the nonstrange $\eta_\NSt (758)$.  But since $\eta_8
(567)$ appears far from the NG massless limit we must ask: how
close is $\eta_0 (947)$ to the chiral-limiting nonvanishing singlet
$\eta$ mass?

To answer this latter question, return to Fig. 1 and the quark
annihilation strength $\beta \approx 0.28$ GeV$^2$ in Eq.~(\ref{eqno7}).
These $\overline q q$ states presumably hadronize into the U$_A$(1) singlet
state (\ref{eta0def}),
for effective squared mass in the SU(3) limit with $\beta$ remaining
unchanged \cite{20}:
\eqnb
	m^2_{\eta _0} = 3 \beta \approx (917 ~\w{MeV})^2~.
	\label{eqno10}
\eqne
\noindent
This latter CL $\eta_0$ mass in (\ref{eqno10}) is only 3\% shy of the exact
chiral-broken $\eta_0 (947)$ mass found in Eq.~(\ref{eqno1}). (Such a 3\% CL 
reduction also holds for the pion decay constant $f_\pi \approx 93$ 
MeV $\to 90$ MeV \cite{22} and for $f_+ (0) = 1 \to 0.97$ \cite{23}, the 
$K$--$\pi$ $K_{l3}$ form factor.)

Our $\eta$--$\eta^\prime$ mixing analysis on the basis of phenomenological
mass inputs thus tells us that the physical $\eta (547)$ is 97\% of the
{\it chiral-broken} NG boson $\eta_8 (567)$.  Also the mixing-induced CL
singlet mass of 917 MeV in (\ref{eqno10}) is 97\% of the chiral-broken
singlet $\eta_0 (947)$ in (\ref{eqno1}), which in turn is 99\% of the
physical $\eta'$ mass $\eta' (958)$. This can be viewed as the 
phenomenological resolution of the U$_A$(1) problem of the masses and
(quasi-)Goldstone boson structure of the observed $\eta(547)$ and
$\eta^\prime(958)$ mesons. Or rather, from a more microscopic standpoint,
the above represents phenomenological constraints that microscopic, 
more or less QCD--based studies of the $\eta$--$\eta^\prime$ 
complex must respect.

{\section{Bound-state SD--BS approach to \mbox{$\eta$--$\eta^\prime$}}
\label{MainSec}

The coupled Schwinger-Dyson (SD) and Bethe-Salpeter (BS)
approach \cite{Roberts0007054}
can be formulated so that it has strong and clear connections with
QCD, the fundamental theory of strong interactions. In this approach, by
solving the SD equation for dressed quark propagators of various flavors,
one explicitly constructs constituent quarks. They in turn build $q\bar q$
meson bound states which are solutions of the BS equation employing the
dressed quark propagator obtained as the solution of the SD equation.
If the SD and BS equations are so coupled in a consistent approximation,
the light pseudoscalar mesons are simultaneously the $q\bar q$ bound states
and the (quasi) Goldstone bosons of dynamical chiral symmetry breaking
(D$\chi$SB). The resulting relativistically covariant bound-state
model (such as the variant of Ref.~\cite{jain93b}) is consistent with
current algebra because it incorporates the correct chiral symmetry behavior
thanks to D$\chi$SB obtained in an, essentially, Nambu--Jona-Lasinio fashion,
but the SD--BS model interaction is less schematic.
In Refs. \cite{KlKe2,jain93b,munczek92,KeBiKl98,KeKl1,KeKl3} for
example, it is combined nonperturbative and perturbative gluon exchange;
the effective propagator function is the sum of the known perturbative
QCD contribution and the modeled nonperturbative component.
For details, we refer to Refs. \cite{KlKe2,jain93b,munczek92,KeBiKl98,KeKl1},
while here we just note that the momentum-dependent dynamically generated
quark mass functions ${\cal M}_f(q^2)$ ({\it i.e.}, the quark propagator
SD solutions for quark flavors $f$) illustrate well
how the coupled SD-BS
approach provides a modern constituent model which is consistent
with perturbative and nonperturbative QCD. For example, the perturbative
QCD part of the gluon propagator leads to the deep Euclidean behaviors
of quark propagators (for all flavors) consistent with the asymptotic
freedom of QCD \cite{KeBiKl98}.
However, what is important in the present paper, is the behavior
of the mass functions ${\cal M}_f(q^2)$ for {\it low} momenta
[$q^2=0$ to $-q^2\approx (400 \, {\rm MeV})^2$], where ${\cal M}_f(q^2)$
(due to D$\chi$SB) has values consistent with typical values
of the constituent mass parameter in constituent quark models.
For the (isosymmetric) $u$- and $d$-quarks, our concrete model choice
\cite{jain93b} gives us ${\cal M}_{u,d}(0)=356$ MeV in the chiral limit
({\it i.e.}, with vanishing ${\widetilde m}_{u,d}$, the explicit chiral symmetry
breaking bare mass term in the quark propagator SD equation, resulting in
vanishing pion mass eigenvalue, $m_\pi=0$, in the BS equation),
and ${\cal M}_{u,d}(0)=375$ MeV [just 5\% above ${\cal M}_{u,d}(0)$ in the
chiral limit] with the bare mass
${\widetilde m}_{u,d}=3.1$ MeV, leading to a realistically light pion,
$m_\pi=140.4$ MeV. Similarly, for the $s$ quark, ${\cal M}_s(0)=610$ MeV.
The simple-minded constituent masses in both {\it NS} and {\it S} sectors, 
$\hat{m}$ and $\hat{m}_s$ employed in Sec. \ref{Preliminaries}, have thus 
close analogues in the coupled
SD--BS approach which explicitly incorporates some crucial features of QCD,
notably D$\chi$SB. Thanks to D$\chi$SB, this dynamical, bound-state approach 
successfully incorporates the partially Goldstone boson structure of the 
mixed $\eta (547)$ and $\eta' (958)$ mesons \cite{KlKe2}. 

Before addressing its mass matrix, let us briefly recall what the SD--BS 
approach revealed \cite{KlKe2,KeKlSc2000} about the mixing angle inferred 
from $\eta, \eta^\prime \to \gamma\gamma$ decays.
The SD--BS approach incorporates the correct chiral symmetry behavior 
thanks to D$\chi$SB and is consistent with current algebra. Therefore, 
and this gives particular weight to the constraints placed on the mixing angle
$\theta$ by the SD-BS results on $\gamma\gamma$ decays of pseudoscalars,
this approach reproduces
(when care is taken to preserve the vector Ward-Takahashi identity of QED)
analytically and exactly the CL pseudoscalar $\to \gamma\gamma$ decay 
amplitudes ({\it e.g.}, $\pi^0\to \gamma\gamma$), which are fixed by the 
Abelian axial anomaly.
(Note that they are otherwise notoriously difficult to reproduce 
in bound-state approaches, as discussed in Ref.~\cite{KeBiKl98}.) 

General and robust considerations in this chirally well-behaved
approach showed \cite{KlKe2} that, unlike the pion case, 
$\eta_8 , \eta_0 \to \gamma \gamma$
(and therefore also their mixtures $\eta, \eta^\prime \to \gamma \gamma$) 
decay amplitudes cannot be given through their respective axial-current
decay constants $f_{\eta_8}, f_{\eta_0}$, and also gave strong bounds on
these amplitudes with respect to the pion decay constant $f_\pi$
({\it i.e.}, w. r. to the $\pi^0\to \gamma \gamma$ amplitude).
All this says that in models relying on quark degrees of
freedom, reasonably accurate reproduction of the
empirical $\eta,\eta^\prime\to \gamma\gamma$ widths is possible 
only for $\theta$-values less negative than $-15^\circ$. For the 
concrete \cite{munczek92,jain93b} model adopted in Ref. \cite{KlKe2}, 
our calculated $\eta,\eta^\prime\to\gamma\gamma$ widths 
fit the data best for $\theta = -12.0^\circ$.

For the very predictive SD-BS approach to be consistent,
the above mixing angle extracted
from $\eta,\eta^\prime\to\gamma\gamma$ widths, should be
close to the angle $\theta$ predicted by diagonalizing the
$\eta$--$\eta^\prime$ mass matrix.  
In this section, it is given in the quark $f\bar f$ basis: 
\begin{equation}
{\hat M}^2 = \mbox{\rm diag} (M_{u\bar{u}}^2,M_{d\bar{d}}^2,M_{s\bar{s}}^2)
+ \beta 
\left( \begin{array}{ccl} 1 & 1 & 1 \\
                          1 & 1 & 1 \\
                          1 & 1 & 1
        \end{array} \right)~.
\label{M2}
\end{equation}
As in Sec. \ref{Preliminaries}, $3\beta$ (called $\lambda_\eta$
in Ref.~\cite{KlKe2}) is the contribution of the gluon axial anomaly
to $m_{\eta_0}^2$, the squared mass of $\eta_0$.
We denote by $M_{f{\bar f}^\prime}$ the masses obtained as eigenvalues 
of the BS equations for $q\bar q$ pseudoscalars with the
flavor content ${f{\bar f}^\prime}$ ($f, f^\prime = u, d, s$).
However, since Ref.~\cite{KlKe2} had to employ a rainbow-ladder approximation 
(albeit the improved one of Ref.~\cite{jain93b}), it could not calculate the 
gluon axial anomaly contribution $3\beta$. It could only avoid the
U$_A$(1)-problem in the $\eta$--$\eta^\prime$ complex by {\it parameterizing} 
$3\beta$, namely that part of the $\eta_0$ mass squared which remains 
nonvanishing in the CL. Because of the rainbow-ladder approximation (which 
does not contain even the simplest annihilation graph -- Fig. 1), the 
$q\bar q$ pseudoscalar masses $M_{f{\bar f}^\prime}$ {\it do not} contain 
any contribution from $3\beta$, unlike the
nonstrange and strange $\eta$ masses $m_{\eta_{NS}}$ [in Eq.~(\ref{eqno9a})]
and $m_{\eta_S}$ [in Eq.~(\ref{eqno9b})], which do, and which must not
be confused with $M_{u\bar u}=M_{d\bar d}$ and $M_{s\bar s}$.
Since the flavor singlet gluon anomaly contribution $3\beta$ does not 
influence the masses $m_\pi$ and $m_K$ of the non-singlet pion and kaon, 
the realistic rainbow-ladder modeling aims directly at
reproducing the empirical values of these masses: 
$M_{u\bar u}=M_{d\bar d}=m_\pi$ and $M_{s\bar d} = m_K$. In contrast, 
the masses of the physical etas, $m_\eta$ and $m_{\eta^\prime}$, must be 
obtained by diagonalizing the $\eta_8$-$\eta_0$ sub-matrix containing 
both $M_{f\bar f}$ and the gluon anomaly contribution to $m_{\eta_0}^2$.

Since the gluon anomaly contribution $3\beta$ vanishes in the large $N_c$ 
limit as $1/N_c$, while all $M_{f{\bar f}^\prime}$ vanish in CL, our $q\bar q$ 
bound-state pseudoscalar mesons behave
in the $N_c\to\infty$ and chiral limits in agreement with QCD and
$\chi$PT ({\it e.g.}, see \cite{G+L}): as the strict CL
is approached for all three flavors, the SU(3) octet pseudoscalars
{\it including} $\eta$ become massless Goldstone bosons, whereas the
chiral-limit-nonvanishing $\eta^\prime$-mass $3\beta$ is of order $1/N_c$ 
since it is purely due to the gluon anomaly. 
If one lets $3\beta \to 0$ (as the gluon anomaly contribution 
behaves for $N_c\to\infty$), then for any quark masses and 
resulting $M_{f\bar f}$ 
masses, the ``ideal" mixing ($\theta=-54.74^\circ$) takes place so that 
$\eta$ consists of $u,d$ quarks only and becomes 
degenerate with $\pi$, whereas $\eta^\prime$ is the pure $s\bar s$ 
pseudoscalar bound state with the mass $M_{s\bar s}$.

In Ref.~\cite{KlKe2}, numerical calculations of the mass matrix were 
performed for the realistic chiral and SU(3) symmetry breaking, 
with the finite quark masses (and thus also the finite BS $q\bar q$ bound-state 
pseudoscalar 
masses $M_{f\bar f}$) fixed by the fit~\cite{jain93b} to static properties 
of many mesons but excluding the $\eta$--$\eta^\prime$ complex. The mixing 
angle which diagonalizes the  $\eta_8$-$\eta_0$ mass matrix thus depended 
in Ref.~\cite{KlKe2} only on the value of the additionally introduced 
``gluon anomaly parameter" $3 \beta$. Its preferred value 
turned out to 
be $3 \beta=1.165$ GeV$^2$=(1079 MeV)$^2$, leading to the mixing 
angle $\theta=-12.7^\circ$ 
[compatible with $\phi=41.84^\circ$ in Eq.~(\ref{eqno8})]
and acceptable $\eta\to \gamma \gamma$ and 
$\eta^\prime\to \gamma\gamma$ decay amplitudes. Also, the $\eta$ mass was 
then fitted to its experimental value, but such a high value of $3\beta$ 
inevitably resulted in a too high  $\eta^\prime$ mass, above 1 GeV.
(Conversely, lowering $3\beta$ aimed to reduce $m_{\eta^\prime}$, 
would push $\theta$ close to $-20^\circ$, making predictions
for $\eta,\eta^\prime \to \gamma\gamma$ intolerably bad.)
However, unlike Eq.~(\ref{eqno6}) in the present paper, it should be 
noted that Ref.~\cite{KlKe2} did not introduce into the mass matrix
the ``strangeness attenuation parameter" $X$ which should suppress 
the nonperturbative quark $f\bar f \to f^\prime {\bar f}^\prime$ 
annihilation amplitude (illustrated by the ``diamond" graph in Fig. 1)
when $f$ or $f^\prime$ are strange.
Ref. \cite{KeKlSc2000} concluded that it was precisely
the lack of the strangeness attenuation factor $X$
that prevented Ref. \cite{KlKe2} from satisfactorily
reproducing the $\eta^\prime$ mass when it successfully did so
with the $\eta$ mass and $\gamma\gamma$ widths.

One can expect that the influence of this suppression should be
substantial, since $X \approx {\hat m}/\hat{m}_s$ should be a reasonable 
estimate of it, and this nonstrange-to-strange {\it constituent} mass
ratio in the considered variant of the SD-BS approach~\cite{KlKe2} is 
not far from $X$ in Eq.~(\ref{eqno7}) and from the mass ratios in Refs.~\cite{17,18,19}, 
and is even closer to the mass ratios in the Refs. \cite{16}.
Namely, two of us found~\cite{KlKe2} it to be around 
${\cal M}_u(0)/{\cal M}_s(0)=0.615 $ if the constituent mass was
defined at the vanishing argument $q^2$ of the momentum-dependent 
SD mass function ${\cal M}_f(q^2)$.

We therefore introduce the suppression parameter $X$ 
the same way as in the {\it NS--S} mass matrix (\ref{eqno6}), 
whereby the mass matrix in the $f\bar f$ basis becomes 
\begin{equation}
{\hat M}^2 = \mbox{\rm diag} (M_{u\bar{u}}^2,M_{d\bar{d}}^2,M_{s\bar{s}}^2)
    + \beta
        \left( \begin{array}{ccl} 1 & 1 & X \\
                                  1 & 1 & X \\
                                  X & X & X^2
        \end{array} \right)~.
\label{M2wX}
\end{equation}
The very accurate isospin symmetry makes the mixing of
the isovector $\pi^0$ and the isoscalar etas negligible
for all our practical purposes. 
Going to a meson basis of $\pi^0$ and etas enables us
therefore to separate the $\pi^0$ and restrict ${\hat M}^2$ to 
the $2\times 2$ subspace of the etas. In the {\it NS--S} basis,
\eqnb
        \pip{
                \begin{array}{ll}
         m_{\eta_{NS}}^2    &  m_{\eta_{S}\eta_{NS}}^2 \\
            m_{\eta_{NS}\eta_{S}}^2 &    m_{\eta_{S}}^2
                \end{array}
        }
     =
        \pip{
                \begin{array}{ll}
      M_{u\bar{u}}^2 + 2 \beta  & \quad \sqrt{2} \beta X \\
        \,  \sqrt{2} \beta X    & M_{s\bar{s}}^2 + \beta X^2
                \end{array}
        }.
        \label{SD-BS-NS-S}
\eqne
To a very good approximation, Eq. (\ref{SD-BS-NS-S}) recovers 
Eq.~(\ref{eqno6}). This is because not only
$m_\pi=M_{u\bar u}=M_{d\bar d}$, but also because
$M_{s\bar s}^2$ differs from $2 m_K^2 - m_\pi^2$ only by a couple
of percent, thanks to the good chiral behavior of the 
masses $M_{f{\bar f}^\prime}$ calculated in SD-BS approach.
(These $M_{f\bar{f}^\prime}^2$ and the CL model values of $f_\pi$ and 
quark condensate, satisfy Gell-Mann-Oakes-Renner relation to first 
order in the explicit chiral symmetry breaking \cite{munczek92}.)
The SD-BS--predicted octet (quasi-)Goldstone masses $M_{f{\bar f}^\prime}$ 
are known to be empirically successful in our concrete model choice
\cite{jain93b}, but the question is whether the SD-BS approach can also 
give some information on the $X$-parameter.  
If we treat {\it both} $3\beta$ and $X$ as free parameters, we can of 
course fit both the $\eta$ mass and the $\eta^\prime$ mass to their 
experimental values. 
For the model parameters as in Ref. \cite{jain93b}
(for these parameters our independent calculation gives 
$m_\pi=M_{u\bar u}=140.4$ MeV and $M_{s\bar s}=721.4$ MeV),
this 
happens at $3\beta=0.753$ GeV$^2$ =(868 MeV)$^2$
and $X=0.835$.  However,
the mixing angle then comes out as $\theta=-17.9^\circ$, which is 
too negative to allow consistency of the empirically found two-photon
decay amplitudes of $\eta$ and $\eta^\prime$, with predictions of
our SD-BS approach for the two-photon decay amplitudes of $\eta_8$ and 
$\eta_0$ \cite{KlKe2}. 

Therefore, and also to avoid introducing another free parameter in 
addition to $3\beta$, we take the path where the dynamical information 
from our SD-BS approach is used to estimate $X$.
Namely, our $\gamma\gamma$ decay amplitudes $T_{f\bar f}$
can be taken as a serious guide for estimating the $X$-parameter 
instead of allowing it to be free. 
We did point out in Sec. \ref{Preliminaries}
that the attempted treatment \cite{20} of the gluon anomaly contribution
through just the ``diamond diagram" contribution to $3\beta$,
indicated that just this partial contribution is quite insufficient.
This limits us to keeping $3\beta$ as a free parameter,
but we can still suppose that this diagram can help us get the
prediction of the strange-nonstrange {\it ratio} of the complete
pertinent amplitudes $f\bar f \to f^\prime {\bar f}^\prime$ as follows.
Our SD-BS modeling in Ref. \cite{KlKe2} employs an infrared-enhanced
gluon propagator \cite{jain93b,KeBiKl98} weighting the integrand strongly
for low gluon momenta squared.
Therefore, in analogy with Eq.~(4.12) of Kogut and
Susskind \cite{4} (see also Refs. \cite{FrankMeissner97,hep-ph9707210}),
we can approximate the Fig. 1 amplitudes $f\bar f \to {\rm 2 gluons}
\to f^\prime {\bar f}^\prime$, {\it i.e.}, the contribution of the quark-gluon
diamond graph to the element $f f^\prime$ of
the $3\times 3$ mass matrix, by the factorized form
\begin{equation}
{\widetilde T}_{f\bar f}(0,0) 
\, {\cal C} \, \, {\widetilde T}_{f^\prime {\bar f}^\prime}(0,0) \, .
\label{factoriz} 
\end{equation}
In Eq. (\ref{factoriz}), the quantity ${\cal C}$ is given by the 
integral over two gluon propagators remaining after factoring out 
${\widetilde T}_{f\bar f}(0,0)$ and 
${\widetilde T}_{f^\prime {\bar f}^\prime}(0,0)$, the respective 
amplitudes for the transition of the $q\bar q$ pseudoscalar bound 
state for the quark flavor $f$ and $f^\prime$ into two vector bosons, 
in this case into two gluons.
The contribution of Fig. 1 is thereby expressed with the help of
the (reduced) amplitudes ${\widetilde T}_{f\bar f}(0,0)$ calculated
in Ref. \cite{KlKe2} for the transition of $q\bar q$ pseudoscalars 
to two real photons ($k^2={k^\prime}^2=0$), while in general 
${\widetilde T}_{f\bar f}(k^2,k^{\prime 2}) \equiv
{T}_{f\bar f}(k^2,k^{\prime 2})/Q_f^2$ are the ``reduced" two-photon
amplitudes obtained by removing the squared charge factors $Q_f^2$
from ${T}_{f\bar f}$, the $\gamma\gamma$ amplitude of the pseudoscalar
$q\bar q$ bound state of the hidden flavor $f {\bar f}$. 
Although ${\cal C}$ is in principle computable, 
all this unfortunately does not amount to determining $\beta, \beta X$ 
and $\beta X^2$ in Eq. (\ref{M2wX}) since the higher (four-gluon, 
six-gluon, ... , etc.) contributions are clearly lacking. We therefore 
must keep the total (light-)quark annihilation strength $\beta$ as a 
free parameter. However, if we assume
that the suppression of the diagrams with the strange quark in a loop
is similar for all of them, Eq.~(\ref{factoriz}) and the ``diamond" 
diagram in  Fig. 1 help us to at least estimate the parameter $X$ as 
$X \approx {\widetilde T}_{s\bar s}(0,0)/{\widetilde T}_{u\bar u}(0,0)$.
This is a natural way to build in the effects of the SU(3) flavor 
symmetry breaking in the $q\bar q$ annihilation graphs.
(Recall that ${\widetilde T}_{s\bar s}(0,0)/{\widetilde T}_{u\bar u}(0,0)
\approx \hat{m}/\hat{m}_s$ to a good approximation \cite{KeKlSc2000}.)

We get $X=0.663$ from the two-photon amplitudes we obtained in the chosen 
SD-BS model \cite{jain93b}. This value of $X$ agrees well with the other 
way of estimating $X$, namely the nonstrange-to-strange constituent mass 
ratio of Refs. \cite{17,18,19}. 
With $X=0.663$, requiring that 
the $2\times 2$ matrix trace, $m_\eta^2 + m_{\eta^\prime}^2$,
be fitted to its empirical value,
fixes the chiral-limiting nonvanishing singlet 
mass squared to $3\beta=0.832$ GeV$^2$=(912 MeV)$^2$, just 0.5\%
below Eq.~(\ref{eqno10}), while $m_{\eta_{NS}}=757.87$ MeV and 
$m_{\eta_S}=801.45$ MeV, practically the same as Eqs. (\ref{eqno9}). 
The resulting mixing angle and $\eta$, $\eta^\prime$ masses are
\begin{equation}
\phi = 41.3^\circ \wsep{.1}{or} \theta = - 13.4^\circ \, \,  ;
\quad m_\eta = 588 \, \mbox{MeV} \, , 
\quad m_{\eta^\prime} = 933 \, \mbox{MeV} \, .
\label{results1}
\end{equation}
These results are for the original parameters of Ref. \cite{KlKe2}.
Reference \cite{KeKlSc2000} also varied the parameters to check the
sensitivity on SD-BS modeling, but the results changed little.

The above results of the SD-BS approach~\cite{KlKe2} are very satisfactory 
since they agree very well with what was found in Sec. \ref{Preliminaries} 
by different methods. They also agree with the UKQCD lattice results 
\cite{UKQCDetas2000} on $\eta$--$\eta^\prime$ mixing. Their calculated 
mixing parameter $x_{ss}$ corresponds to our $\beta X^2$, and their mixing 
parameters $x_{nn}$ and $x_{ns}$ ($n=u,d$), corresponding respectively
to our $\beta$ and $\beta X$, are aimed to obey $x_{nn} \approx 2 x_{ss}$
and $x_{ns}^2 \approx x_{nn} x_{ss}$. UKQCD prefers \cite{UKQCDetas2000} 
$x_{ss} = 0.13$ GeV$^2$, $x_{nn} = 0.292$ GeV$^2$ and $x_{ns} = 0.218$
GeV$^2$. This, together with their preferred {\it input} values 
$M_{n\bar n}=0.137$ GeV and $M_{s\bar s}=0.695$ GeV, give the 
{\it NS--S} mass matrix (\ref{SD-BS-NS-S}) with elements reasonably 
close to ours, resulting in a rather close mixing angle, 
$\theta = - 10.2^\circ$.

\section{Conclusion}
We have shown that the treatment of the $\eta$--$\eta^\prime$ complex in the 
SD--BS approach \cite{KlKe2} is sensible in spite of employing the ladder 
approximation. This is confirmed especially by Ref. \cite{KeKlSc2000} which 
showed its connection and robust agreement with the phenomenological studies 
of the $\eta$--$\eta^\prime$ complex. It is therefore desirable to extend the 
SD--BS studies of the $\eta$--$\eta^\prime$ mass matrix to finite temperatures.
Usually, one has neglected all temperature dependences in the mass matrix, 
except the one of the gluon anomaly contribution $3\beta$ which is assumed 
very strong, which is appropriate if the U$_A$(1) symmetry breaking is due to 
instantons \cite{PisarskiWilczek84,AlkofAmundReinh89,Kapusta+al96,HuangWang96}.
However, rather strong topological arguments of Kogut {\it et al.} 
\cite{Kogut+al98} that the U$_A$(1) symmetry is not restored at critical 
(but only at a higher, possibly infinite) $T$, motivates also the scenario 
where $3\beta(T)\approx const$, while other entries in the mass matrix carry 
the temperature dependence. The inclusion of their $T$--dependence is needed 
also because the scenario with the instanton--induced, strongly $T$--dependent 
$\beta$ should be carefully re-examined, since it has lead to contradicting 
conclusions:
the depletion of $\eta^\prime$ production in Ref. \cite{AlkofAmundReinh89}, 
but $\eta^\prime$--enhancement in Ref. \cite{Kapusta+al96}.

The temperature dependence of $m_\pi = M_{u\bar u} =  M_{d\bar d}$, 
${\cal M}_{u,d}(q^2)$, $f_\pi$ and 
$\langle{u\bar u}\rangle (= \langle{d\bar d}\rangle)$, was already studied 
in various SD--BS models \cite{Maris+alNT0001064,Blaschke+alNT0002024}, so 
that the extension \cite{Blaschke+al01} to the $T$--dependence of 
the remaining needed ingredients, $M_{s\bar s}$, ${\cal M}_{s}(q^2)$, 
$f_{s\bar s}$ and $\langle{s\bar s}\rangle$, should be straightforward.

\vskip 5mm

\noindent {\bf Acknowledgments:} 
D. Kl. thanks thanks D. Blaschke, S. Schmidt and G. Burau, the organizers
of the workshop ``Quark Matter in Astro- and Particle Physics" 
(27.-29. November 2000, Rostock, Germany) for their hospitality and 
for the support which made his participation possible.


\end{document}